\documentclass[%
 reprint,
superscriptaddress,
 amsmath,amssymb,
 aps,
pra,
]{revtex4-1}

\usepackage{graphicx}
\usepackage{dcolumn}
\usepackage{bm}
\usepackage{xcolor}
\usepackage{siunitx}
\begin{document}

\preprint{APS/123-QED}

\title{Revealing Laser and Electron Beam Evolution in 10-GeV-class Laser–Plasma Accelerators}

\author{H.~Tang}
\affiliation{Lawrence Berkeley National Laboratory, Berkeley, California 94720, USA}%
\author{A.~Picksley}%
\affiliation{Lawrence Berkeley National Laboratory, Berkeley, California 94720, USA}%
\author{C.~Benedetti}
\affiliation{Lawrence Berkeley National Laboratory, Berkeley, California 94720, USA}%
\author{R.~Li}
\affiliation{Lawrence Berkeley National Laboratory, Berkeley, California 94720, USA}%
\affiliation{University of California, Berkeley, California 94720, USA}%
\author{H.~E.~Tsai}
\affiliation{Lawrence Berkeley National Laboratory, Berkeley, California 94720, USA}%
\author{T.~Mandal}
\affiliation{Lawrence Berkeley National Laboratory, Berkeley, California 94720, USA}%
\author{E.~Park}
\affiliation{Lawrence Berkeley National Laboratory, Berkeley, California 94720, USA}%
\affiliation{University of California, Berkeley, California 94720, USA}%
\author{K.~Nakamura}
\affiliation{Lawrence Berkeley National Laboratory, Berkeley, California 94720, USA}%
\author{J.~Stackhouse}
\affiliation{Lawrence Berkeley National Laboratory, Berkeley, California 94720, USA}%
\affiliation{University of California, Berkeley, California 94720, USA}%
\author{D.~Terzani}
\affiliation{Lawrence Berkeley National Laboratory, Berkeley, California 94720, USA}%
\author{C.~B.~Schroeder}
\affiliation{Lawrence Berkeley National Laboratory, Berkeley, California 94720, USA}%
\affiliation{University of California, Berkeley, California 94720, USA}%
\author{J.~van~Tilborg}
\affiliation{Lawrence Berkeley National Laboratory, Berkeley, California 94720, USA}%
\author{J.~Osterhoff}
\affiliation{Lawrence Berkeley National Laboratory, Berkeley, California 94720, USA}%
\author{C.~G.~R.~Geddes}
\affiliation{Lawrence Berkeley National Laboratory, Berkeley, California 94720, USA}%
\author{A.~J.~Gonsalves}
\email{email: ajgonsalves@lbl.gov}
\affiliation{Lawrence Berkeley National Laboratory, Berkeley, California 94720, USA}%

\date{\today}

\begin{abstract}
Guiding relativistically intense laser pulses in low-density plasmas enables extended acceleration lengths in laser–plasma accelerators (LPAs), allowing for the production of multi-GeV electron beams. Quantitative interpretation of such experiments is often limited by substantial uncertainties in key plasma parameters, particularly the transverse density profile of hydrodynamic optically field-ionized channels. Distinct plasma density distributions can produce similar terminal beam energies, complicating efforts to infer the underlying interaction physics from measurements at the accelerator exit alone. By combining longitudinally resolved electron beam diagnostics with independent measurements of laser spectral evolution in a 10 GeV LPA, we establish a multi-observable constraint on plasma density profiles. Once plasma downramps are taken into account,  excellent agreement is observed with simulation over the entire accelerator length for two plasma channel sizes. The validated simulations indicate that extending the accelerator length to 65 cm would increase the electron beam energy to 15 GeV. They also point the way to achieving $\sim$20 GeV electron beams in $\sim \SI{70}{cm}$ via linear matching using the same 24 J laser energy.

\end{abstract}

\maketitle

\section{\label{sec:level1}Introduction}

Laser–plasma accelerators (LPAs) have attracted significant attention as a pathway toward compact high-energy accelerators due to their ability to sustain accelerating gradients exceeding 100 GeV/m \cite{tajima1979laser, esarey2002overview, esarey2009physics}. With the development of petawatt-class laser facilities, laser wakefield acceleration has enabled the production of multi-GeV electron beams over centimeter-scale distances \cite{wang2013quasi, leemans2014multi, gonsalves2019petawatt, miao2022multi, aniculaesei2024acceleration, picksley2024prl, rockafellow2025high}. These advances make LPAs promising candidates for applications including next-generation linear colliders \cite{schroeder2010physics, schroeder2023linear}, compact x-ray free-electron lasers \cite{wang2021free, labat2023seeded, barber2025greater}, and other high-brightness photon sources \cite{geddes2015compact, albert2016applications}.

A preformed plasma channel, which acts like a gradient-index fiber for high-intensity laser pulses, can increase electron bunch energy by increasing the length over which the laser remains intense. Such channels are formed with a radial plasma electron density profile $n(r)$ with axial minimum such that the refractive index is peaked. Plasma channels formed by the hydrodynamic expansion of optical field-ionized (HOFI) plasmas \cite{lemos2013plasma, shalloo2018hydrodynamic, shalloo2019low, smartsev2019axiparabola, morozov2018ionization, picksley2020meter, feder2020self} can provide plasma channels with low density and compatibility with controlled injection techniques \cite{oubrerie2022controlled, picksley2023all}, enabling the generation of high-quality multi-GeV electron beams \cite{picksley2024prl, lahaye2025quasimonoenergetic}. 

In a typical channel-guided LPA experiment, optical guiding and electron acceleration are assessed by measuring electron spectra \cite{shrock2024guided} or laser properties and electron spectra \cite{leemans2014multi, gonsalves2019petawatt, picksley2023all, picksley2024prl} at the exit of the plasma channel. The propagation of the laser pulse and details of the interaction are inferred from simulations. We recently measured guided laser mode and optical spectrum as a function of propagation distance in a 10-GeV-class, channel guided LPA \cite{picksley2024prl}, allowing for direct observation of laser coupling into high-order channel modes and their energy loss through mode-filtering, followed by quasi-matched propagation of the fundamental channel mode, and non-linear depletion of laser energy to the wakefield. 

In all LPA experiments, there are uncertainties in the measured input parameters. For HOFI-based LPA experiments to date \cite{oubrerie2022controlled, miao2022multi, picksley2023all, shrock2024guided, picksley2024prl, rockafellow2025high}, errors in the measurement of the transverse plasma channel density distribution $n(r)$ are dominant. When simulating an experiment, the density profile can be varied within the measurement uncertainty to achieve agreement with the measured electron or electron and laser parameters at the output of the accelerator. Here we show that even when measuring both laser and electron beam properties at the output of the 40-cm-long accelerator, simulations with different density profiles can match the experimentally measured laser redshift and maximum beam energy of $\SI{8}{GeV}$. Measuring the properties of both as a function of plasma length removes this degeneracy, allowing for better physical understanding of the interaction. This in turn shows the path to maximizing the electron beam energy and laser-to-beam transfer efficiency. By requiring simultaneous agreement between the measured laser spectral evolution and the electron beam  energy evolution along the accelerator, we constrain the channel parameters to an on-axis density of $3\times10^{16}$ cm$^{-3}$ and a channel radius of \SI{29}{\micro m}. We show that the laser mode size in simulation and experiment only match when the density downramp that exists for each plasma length is taken into account. With this match of all observables we confirmed that the channel radius was smaller than that for matched guiding. Guided by the validated simulations, we increased the channel-forming laser intensity to increase plasma channel size, which improved guiding and increased maximum electron energy to $\SI{10}{GeV}$. Simulations further show that by increasing plasma length to 60 cm, the beam energy could be increased by another $\SI{50}{\%}$.

\section{\label{sec:level2}Experimental Setup}

Experiments were performed using the Ti:sapphire BELLA petawatt laser system ($\lambda_0 = 815$~nm, repetition rate up to 1~Hz) 
\cite{nakamura2017diagnostics}, which provides two independently controlled beamlines. The channel-forming beamline (derived from the BELLA second beamline capability \cite{picksley2022commissioning, turner2022strong}) contained $\SI{1.0(1)}{J}$, and was focused by an axicon lens to produce a line focus of $\gtrsim \SI{50}{cm}$ and full-width-half-maximum (FWHM) width $\approx \SI{14}{\micro m}$.  Error bars on laser parameters correspond to measurement uncertainty. The pulse compression could be adjusted to vary the peak intensity, and hence the initial plasma column radius. For this work, two peak intensities were employed for channel formation: $I_{\mathrm{ch}} = \SI{3.2(4)e14}{W.cm^{-2}}$ and $I_{\mathrm{ch}} = \SI{1.1(1)e15}{W.cm^{-2}}$, where $I_{\mathrm{ch}}$ is the peak intensity averaged over the line focus. 

\begin{figure}[t]
    \centering
    \hspace{-0.8cm}
    \includegraphics[width=1.0\linewidth]{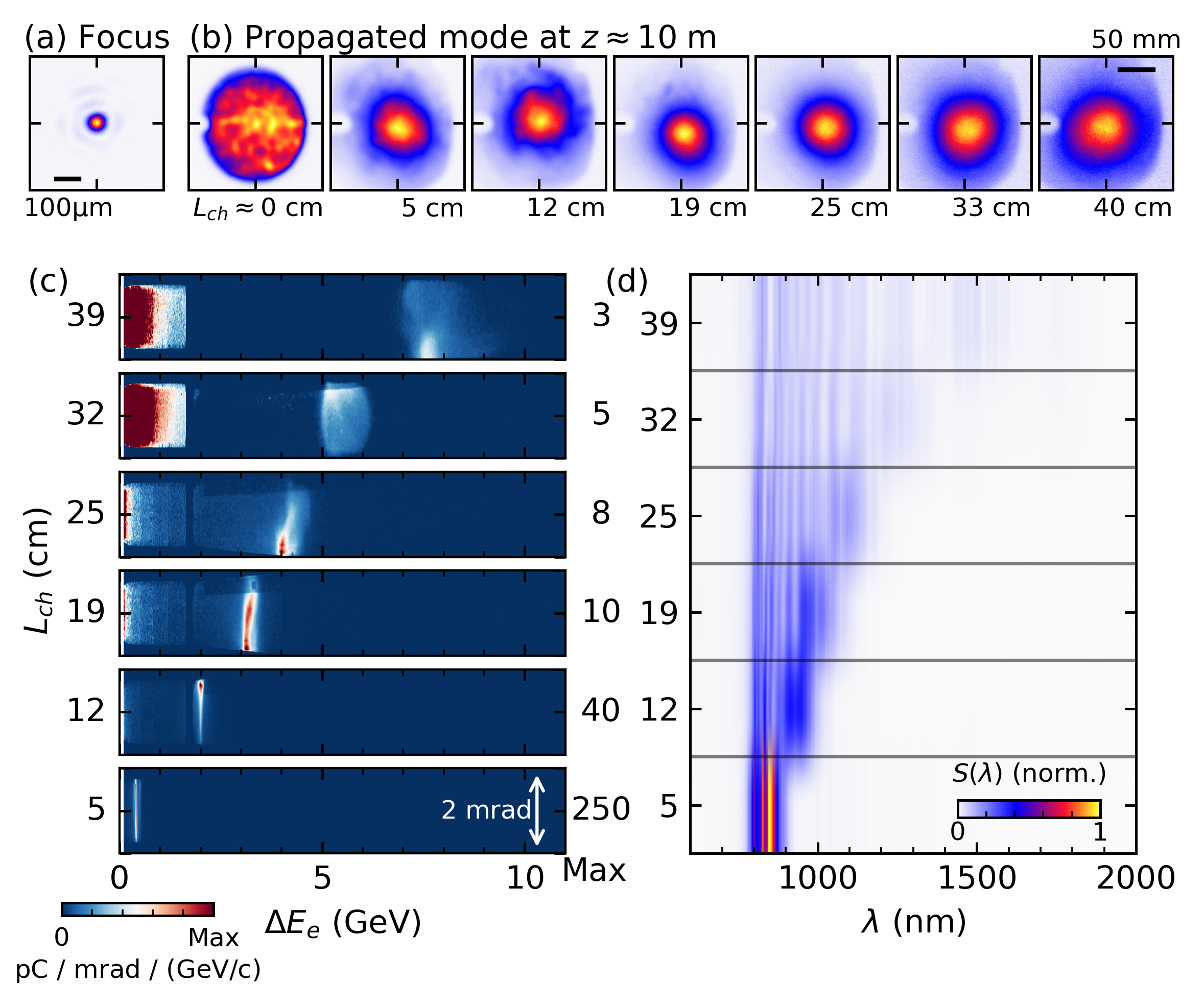}
    \caption{ Evolution of accelerator parameters as a function of channel length for drive laser energy $\mathcal{E}=24~\rm{J}$, $\Delta t=6~\rm{ns}$ and $1\%$ nitrogen dopant located in the first \SI{7.5}{cm}. (a) The re-imaged vacuum mode input into the channel, (b) the propagated optical mode $z \approx \SI{10}{m}$ downstream from the plasma channel exit, (c) the electron spectrum with the highest energy at each $L_{\rm ch}$ and (d) the averaged optical spectrum over 10 shots.
    }
    \label{fig:fig1}
\end{figure}

The wakefield driver (delivered by the original beamline \cite{nakamura2017diagnostics}) contained pulse energy $\mathcal{E} = (24.3 \pm 0.4)$~J and was compressed to FWHM duration $\tau = (38.5 \pm 2.4)$fs. It was focused by a $f=\SI{13.5}{m}$ off-axis paraboloid to a vacuum spot radius of $w_0 = \SI{50.5(15)}{\micro m}$, corresponding to a peak intensity of $I_{\mathrm{0}} = \SI{1.2(1)e19}{W.cm^{-2}}$ and normalized vector potential $a_0 \approx 2.4$, where $a_0 \approx 0.85\,\lambda_0 [\SI{}{\micro m}] \sqrt{I_0[10^{18}\mathrm{W/cm^2}]}$. The focal profile and vacuum propagated mode of the wakefield driver $z\approx \SI{10}{m}$ downstream of focus are shown in Figs.~\ref{fig:fig1}(a) and (b), respectively. The collimated beam profile of the amplified laser is super-Gaussian, $I_\mathrm{col}(r) \propto \exp[-(r/R)^N]$ with $R\simeq \SI{95}{mm}$ the characteristic radius, and $N \approx 10$, typical of current high-power femtosecond laser systems based on bulk-crystal, Ti:Sapphire. This causes the resulting focal profile (of the approximate form $I(r) \sim [J_1(\beta r) / (\beta r)]^2$, where $\beta$ is a scale factor) to exhibit rings outside of the central maximum. 

A 40-cm-long gas jet \cite{miao2022multi, picksley2024prl, li2025longitudinal} positioned 12~mm below the laser axis was employed as the plasma source. It was supplied by 16 independently controlled gas lines placed equally along the nozzle; the first 7.5~cm of the jet contained a mixture of hydrogen with $1\%$ nitrogen dopant for ionization injection \cite{Pak2010, McGuffey2010, Chen2012}, while the remainder of the jet was filled with pure hydrogen for acceleration. It was possible to vary the acceleration length by partially blocking the jet \cite{picksley2024prl}. 

The delay $\Delta t$ between the two pulses was set to be either $\SI{5}{ns}$ or $\SI{6}{ns}$ to allow sufficient time for hydrodynamic expansion to form the plasma channel. At this time, hydrodynamic expansion of the optical-field-ionized structure has slowed significantly, hence no significant difference to laser guiding was observed between these delays. For these conditions, the plasma channel in which the drive laser was guided $n(r)=n_\mathrm{e}^\mathrm{ax}(r)+n_{\mathrm{n}}(r)$ was measured using two-color transverse interferometry \cite{gonsalves2007transverse, point2014two, feder2020self}(reported in Ref.~\cite{picksley2024prl}) to obtain the electron density $n_\mathrm{e}^\mathrm{ax}$ and neutral density $n_{\mathrm{n}}$ before the arrival of the drive pulse. However, the low signal level of transverse interferometry means that averaging over many hundreds of shots is required.  For cost reasons, this typically means that measurements are taken on smaller laser systems, and the inferred channel profile is subject to significant uncertainty. The low signal level limits the measurement precision, while recovery of the transverse density profile from a single beam requires an Abel inversion, which is sensitive to noise and deviations from cylindrical symmetry. Additional uncertainties also arise from the channel-forming laser intensity, including propagation effects of the channel-forming pulse \cite{miao2024benchmarking}, and from local fluctuations in the gas-jet density. We therefore estimate the uncertainty of the plasma channel parameters axial density $n_0$ and matched spot-size $r_{\rm ch}$ to be $n_0 \approx 1^{+1.0}_{-0.9} \times 10^{17} \; \mathrm{cm^{-3}}$ and $r_{\rm ch} \approx \SI{36(7)}{\micro m}$, respectively. Here, we define $r_{\rm ch}$ as the $1/e^2$ size of the Gaussian that would be matched into the plasma channel.

Diagnostics for the wakefield driver, and accelerated electron beam have been described previously \cite{leemans2014multi, nakamura2017diagnostics, gonsalves2019petawatt, picksley2024prl}. For this work, the input and guided mode of the drive laser could be imaged over a range of $\sim \SI{60}{cm}$, and was also imaged at a plane $\approx \SI{10}{m}$ downstream of the channel exit. The propagated downstream mode provides information about the channel properties regardless of the $z$-position of the guided mode waist, making it useful to directly compare guided mode properties at different plasma channel lengths. The optical spectrum of the drive was measured using fiber-based spectrometers covering the range $\SI{400}{nm} \lesssim \lambda \lesssim \SI{2200}{nm}$. Electron beam diagnostics consisted of a phosphor screen for transverse beam imaging, a TurboICT for charge measurement \cite{nakamura2016pico}, and a 2.5 m magnetic spectrometer for energy spectrum measurements \cite{leemans2014multi}.

\section{\label{sec:level3}Results}

Figure~\ref{fig:fig1} shows (b) the propagated output mode of the laser $z \approx \SI{10}{m}$ downstream of the plasma channel, (c) the electron spectra, and (d) the optical spectrum as a function of plasma channel length for $\Delta t \approx \SI{6}{ns}$, and a $\SI{1}{\%}$ nitrogen dopant located in the region \SIrange{0}{7.5}{cm}. Several factors affected the shot-to-shot fluctuations in the measured output properties. For each jet length, we selected $\sim 15$ shots where the high energy component of the electron beam was best aligned to the magnetic spectrometer. For (b) and (c), single examples are shown with the highest energy spectra, while (d) shows the average over approximately 10 shots. Quasi-monoenergetic electron bunches were observed at all channel lengths. At $L_{\rm ch} \approx \SI{5.4}{cm}$, the electron energy gain was $ \SI{0.42(3)}{GeV}$ with a measured charge of approximately $\SI{20}{pC}$. As the acceleration length increased, the maximum beam energy, $\Delta E_\mathrm{max}$, defined as the energy threshold above which the integrated charge contained in the beam is $\SI{1}{pC}$ grew approximately linearly with channel length $L_{\rm ch}$. The averaged maximum electron energies for $L_{\rm ch} \approx 12, 19, 25, 32,$ and $39$~cm were $2.07\pm0.10$, $3.18\pm0.28$, $3.96\pm0.54$, $5.4\pm0.88$, and $7.81\pm0.84$~GeV, respectively. The measured average acceleration gradient was $\sim \SI{18}{GV.m^{-1}}$. 

The longitudinally resolved redshift in Fig.~\ref{fig:fig1}(d) demonstrates a similar trend. Gradual depletion of laser energy to the wake is observed through gradual red-shifting of the optical spectrum. At $L_{\rm ch} \approx \SI{40}{cm}$, the redshifted wavelength (defined as the wavelength at which the cumulative spectral intensity reaches the 80th percentile) had increased from $\lambda_\mathrm{R}(z\approx0) = \SI{845(5)}{nm}$ to $\lambda_\mathrm{R} = \SI{1410(30)}{nm}$. The guided mode of the laser, propagated $\approx \SI{10}{m}$ downstream of the detector evolved from a super-Gaussian structure to an approximately Gaussian structure over the first few cm of the plasma channel. This indicated rapid filtering of higher-order modes in the first $z \approx \SI{5}{cm}$ of propagation \cite{picksley2024prl}. As $L_{\rm ch}$ increased, the laser remained approximately Gaussian with gradual increase in measured size. 

\subsection{\label{sec:sim}Comparison with simulations}

To fully understand the LPA, we compared the experimental results in Fig.~\ref{fig:fig1} with particle-in-cell (PIC) simulations performed using the quasi-static mode of the ponderomotive, cylindrical code INF\&RNO \cite{benedetti2010efficient, benedetti2017accurate}, part of the \emph{Beam, Plasma \& Accelerator Simulation Toolkit} (BLAST) \cite{blast_website}. For these simulations, the drive laser was initialized using the measured energy, temporal profile, and transverse profile (which was recovered using a Gerchberg-Saxton algorithm). The plasma channel took the form of the measured channel $n(r) = n_\mathrm{e}^\mathrm{ax}(r) + n_\mathrm{n}(r)$ \cite{picksley2024prl}, but the matched radius $r_{\rm ch}$ and on-axis density $n_0$ were varied within the error on the measurement (described in Sec.~\ref{sec:level2}) while retaining the measured radial form. The longitudinal profile of the gas jet consisted of a 1 cm up-ramp followed by a constant-density plateau of length $L_\mathrm{ch}$, and a $\sim \SI{1}{cm}$ down-ramp, consistent with plasma fluorescence measurements. 

\begin{figure}[t]
    \centering
    \includegraphics[width=1\linewidth]{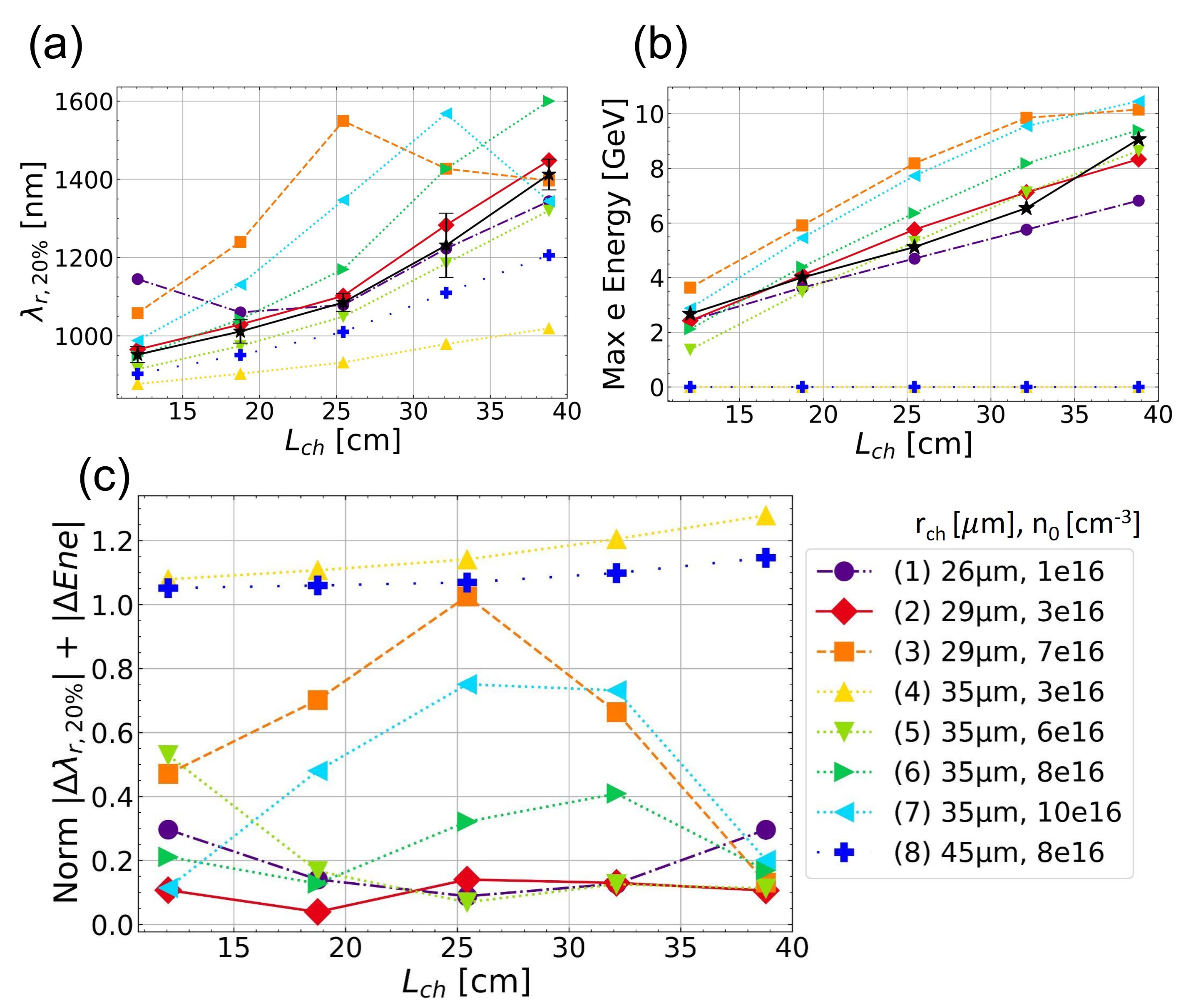}
    \caption{Comparison between experiment (black) and simulations (colors) using a variety of channel inputs within the uncertainty of the measured electron density profile. 
    (a) Laser redshift $\lambda_{r,20\%}$, (b) maximum electron energy, (c) normalized deviations of each simulation from the measured quantities $|\Delta \lambda_{r,10\%}| + |\Delta E_e|$ used to rank simulation inputs.
    }
    \label{fig:fig2}
\end{figure}

We performed simulations for several combinations of $(r_{\rm ch}, n_0)$, and compared the output to two scalar metrics along the accelerator. Figure~\ref{fig:fig2}(a) shows the redshifted wavelength $\lambda_\mathrm{R}$ as a function of $L_\mathrm{ch}$ for eight of those combinations. Since $\lambda_\mathrm{R}$ is a non-linear function of the plasma density, laser field strength, and propagation length, redshift alone does not uniquely determine the transverse density profile. Multiple parameter combinations reproduce the experimental redshift: $(r_{\rm ch}, n_0) = ( \SI{26}{\micro m}, \SI{1e16}{cm^{-3}} ), ( \SI{29}{\micro m}, \SI{3e16}{cm^{-3}} ), ( \SI{35}{\micro m}, \SI{6e16}{cm^{-3}} )$. We therefore additionally compare the maximum electron beam energy (shown in Fig.~\ref{fig:fig2}(b)). Two parameter sets reproduce the measured energy scaling. However, one case, $(r_{\rm ch}, n_0) = ( \SI{35}{\micro m}, \SI{6e16}{cm^{-3}} )$, yields a simulated injected charge of only 0.2~pC, inconsistent with experiment. 
To quantify agreement further, we define a combined metric $|\Delta \lambda_r| + |\Delta E_e|$ based on normalized deviations in redshift and beam energy between simulations and measurement. As shown in Fig.~\ref{fig:fig2}(c), the parameter set $(r_{\rm ch}, n_0) = ( \SI{29}{\micro m}, \SI{3e16}{cm^{-3}} )$ minimizes this metric and simultaneously reproduces the measured redshift, energy evolution, and beam charge. Hence, by requiring agreement with both laser and electron observables, the transverse density profile is constrained.

\begin{figure}[t]
    \centering
    \includegraphics[width=1\linewidth]{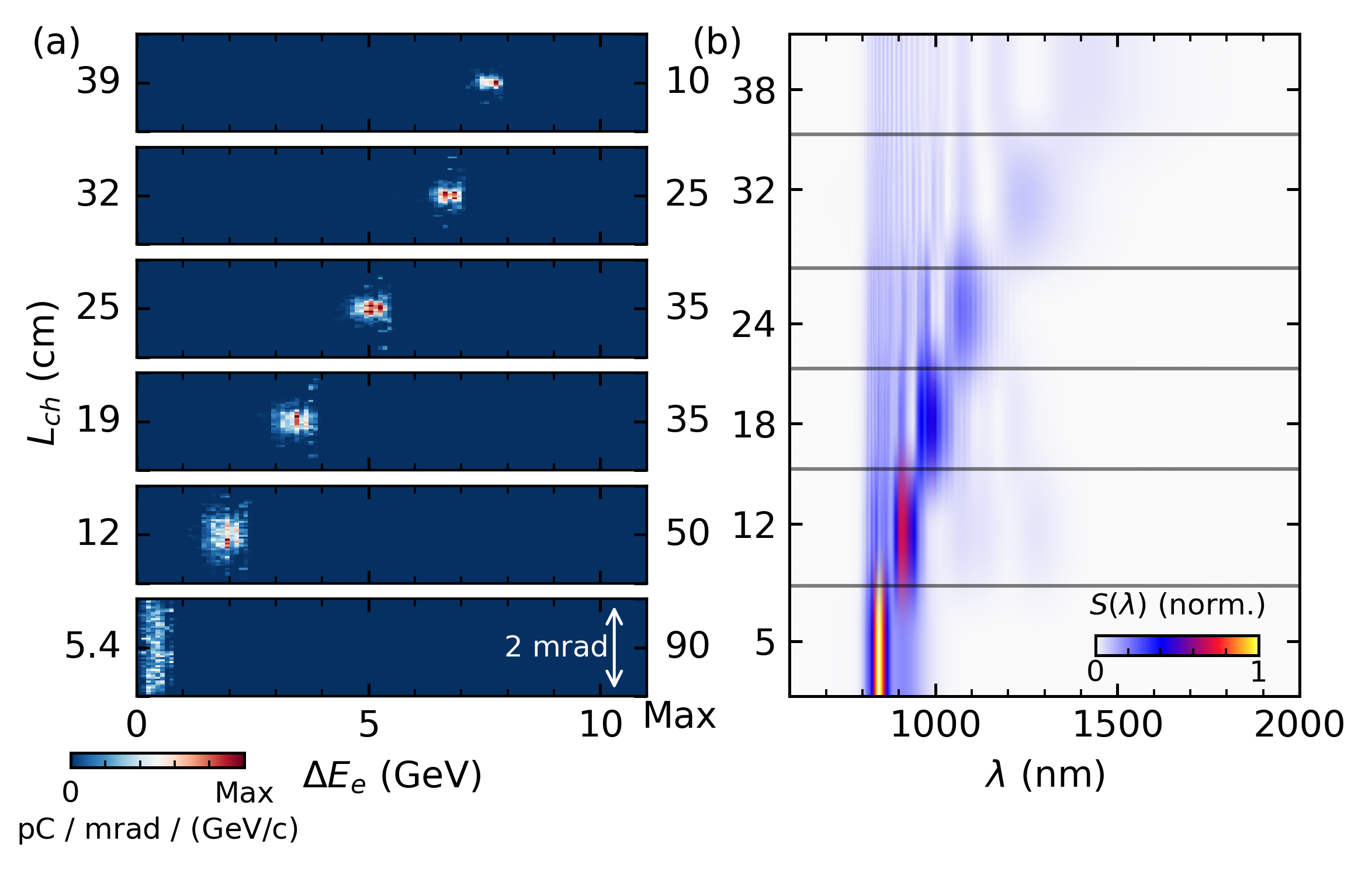}
    \caption{
    Simulated evolution of (a) accelerated electron bunch spectrum, and (b) optical spectrum as a function of channel length for the conditions of Fig.~\ref{fig:fig1}.
    }
    \label{fig:fig3}
\end{figure}

As an additional check of the constrained parameters, Fig.~\ref{fig:fig3} shows (a) the electron beam evolution and (b) the simulated laser spectra, allowing direct comparison of simulation with Fig.~\ref{fig:fig1}(c, d) for more than just the scalar quantities. Excellent agreement is observed. 
Under these conditions, electron injection occurs after $z \sim \SI{3}{cm}$, and by $L_{\rm ch}=38.8$ cm the simulated beam reaches a mean energy of 7.6 GeV with 5.89 pC charge and 0.4 GeV FWHM energy spread. The calculated accelerating gradient was approximately $\approx \SI{20}{GV.m^{-1}}$, consistent with experiment. 

\begin{figure}[t]
    \centering
    \includegraphics[width=1\linewidth]{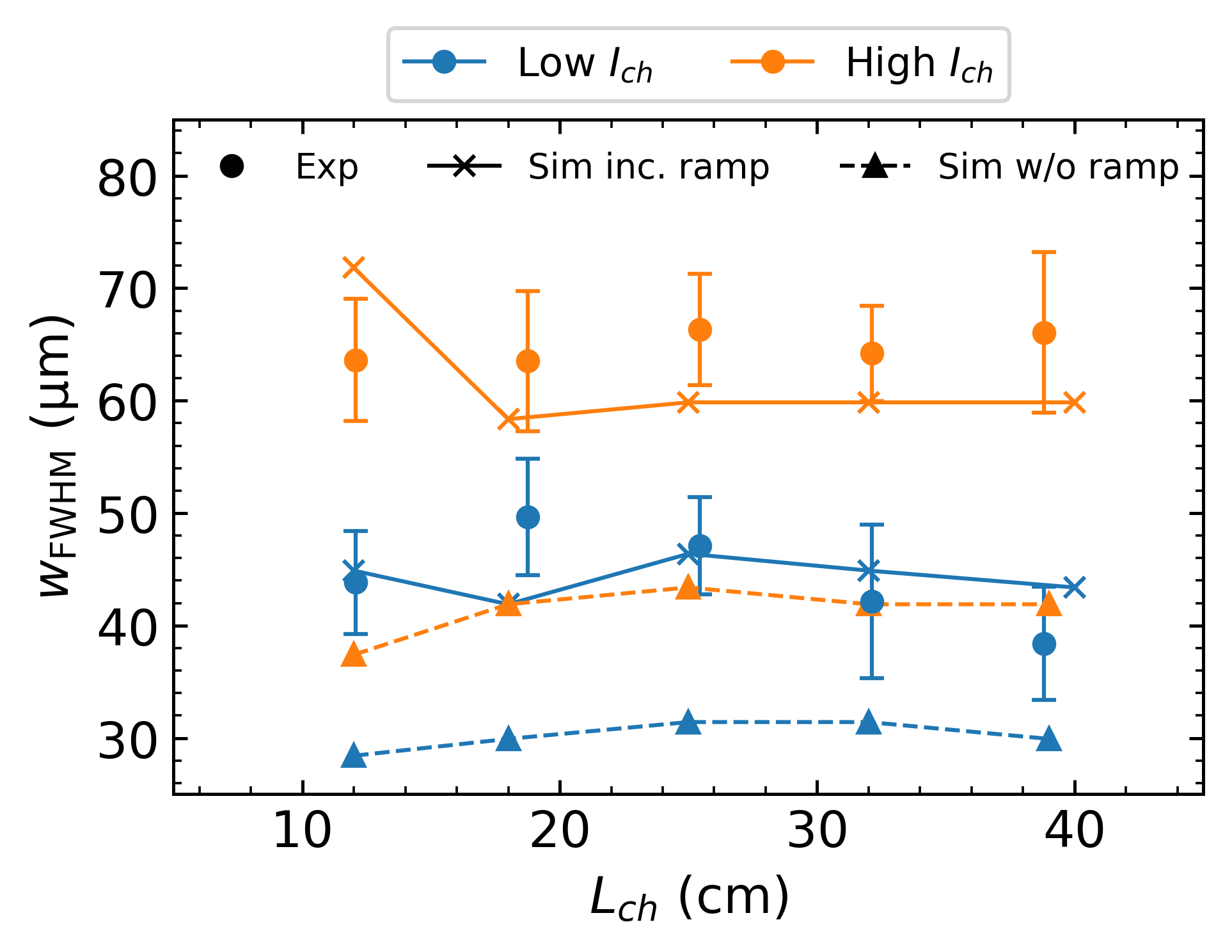}
    \caption{ Laser full-width-at-half-maximum (FWHM) spot-size as a function of $L_\mathrm{ch}$. Markers indicate the inferred size from the propagated mode at $z \approx \SI{10}{m}$ (assuming vacuum propagation) in experiment, averaged over well aligned shots. Dashed lines indicate the simulated FWHM of the laser pulse in the plasma channel, and solid lines indicate simulated FWHM of the laser pulse including the down-ramp at the end of the plasma channel. Two cases are shown, (blue) $I_{\mathrm{ch}} = \SI{3.2(4)e14}{W.cm^{-2}}$  (conditions of Fig.~\ref{fig:fig1}), and (orange) $I_{\mathrm{ch}} = \SI{1.1(1)e15}{W.cm^{-2}}$ (conditions of Fig.~\ref{fig:fig5}). The calculated and inferred mode sizes match only when the down-ramp is included.}
    
    \label{fig:fig4}
\end{figure}

Furthermore, if the simulation accurately captures the plasma channel properties and laser propagation physics, the predicted laser mode evolution should also agree with the measurements. Figure~\ref{fig:fig4} (blue line) compares simulated and measured laser full-width-at-half-maximum (FWHM) spot-size as a function of $L_{\rm ch}$. 
The experimental spot-size was measured approximately $z \approx \SI{10}{m}$ downstream of the interaction (averaged over approximately 10 shots) and back-propagated to the plasma exit assuming vacuum diffraction with perfect phase. 
Triangles come from a single simulation of a 40-cm-long structure and therefore represent the size of the laser mode \textit{inside} the plasma channel. The measurements show a larger mode at each $L_\mathrm{ch}$. This can be explained by laser mode evolution along the 1-cm-long plasma density downramp. The laser mode measurements are not a direct measurements of the laser spot size inside the accelerator. The laser spot-size can change in the density ramp at the exit of the plasma channel since the length of the ramp is comparable with the Rayleigh range of the laser.
When the gas density ramp was accounted for at each channel length (crosses in Fig.~\ref{fig:fig4}) with separate simulations for each $L_\mathrm{ch}$, the simulated and measured spot sizes agree within error. 
We therefore conclude that the calculated values from the case excluding the density ramp provide a good estimate of the true in-plasma laser size (i.e., the matched spot-size of the channel), which is approximately $\SI{15}{\micro m}$ smaller than the value inferred from vacuum back-propagation. We note that redshifting and electron beam energy are relatively unaffected by this exit ramp.

\subsection{\label{sec:level4b}Optimizing Electron Beam Energy}

With increased understanding over the characteristics of the laser and electron beam evolution along the accelerator, we can optimize its performance. 
Increasing the gas pressure could have resulted in increased electron beam energy in the 40~cm-long jet, but was not possible due to limited vacuum pump capacity.
For the laser input shown in Fig.~\ref{fig:fig1}(c), the channel size that maximizes coupling of laser energy into the lowest order mode of the plasma channel is $r_{\rm ch} \approx \SI{47}{\micro m}$, larger than the inferred $r_{\rm ch} \approx \SI{29}{\micro m}$. It was possible to alter $r_{\rm ch}$ by varying the channel-forming laser intensity $I_{\rm ch}$, or $\Delta t$. At times $\Delta t \gtrsim \SI{4}{ns}$, the velocity of the radially propagating shock has reduced significantly, making $I_{\rm ch}$ a better method for altering $r_{\rm ch}$. Hydrodynamic simulations \cite{cook2025hydrodynamic} indicate that increasing the channel-forming laser intensity $I_{\rm ch}$ enables the formation of a wider plasma channel since the initial radius of the plasma cylinder is increased, which in turn leads to a wider shock radius for a fixed $\Delta t$.

\begin{figure}[t]
    \centering
    \includegraphics[width=1\linewidth]{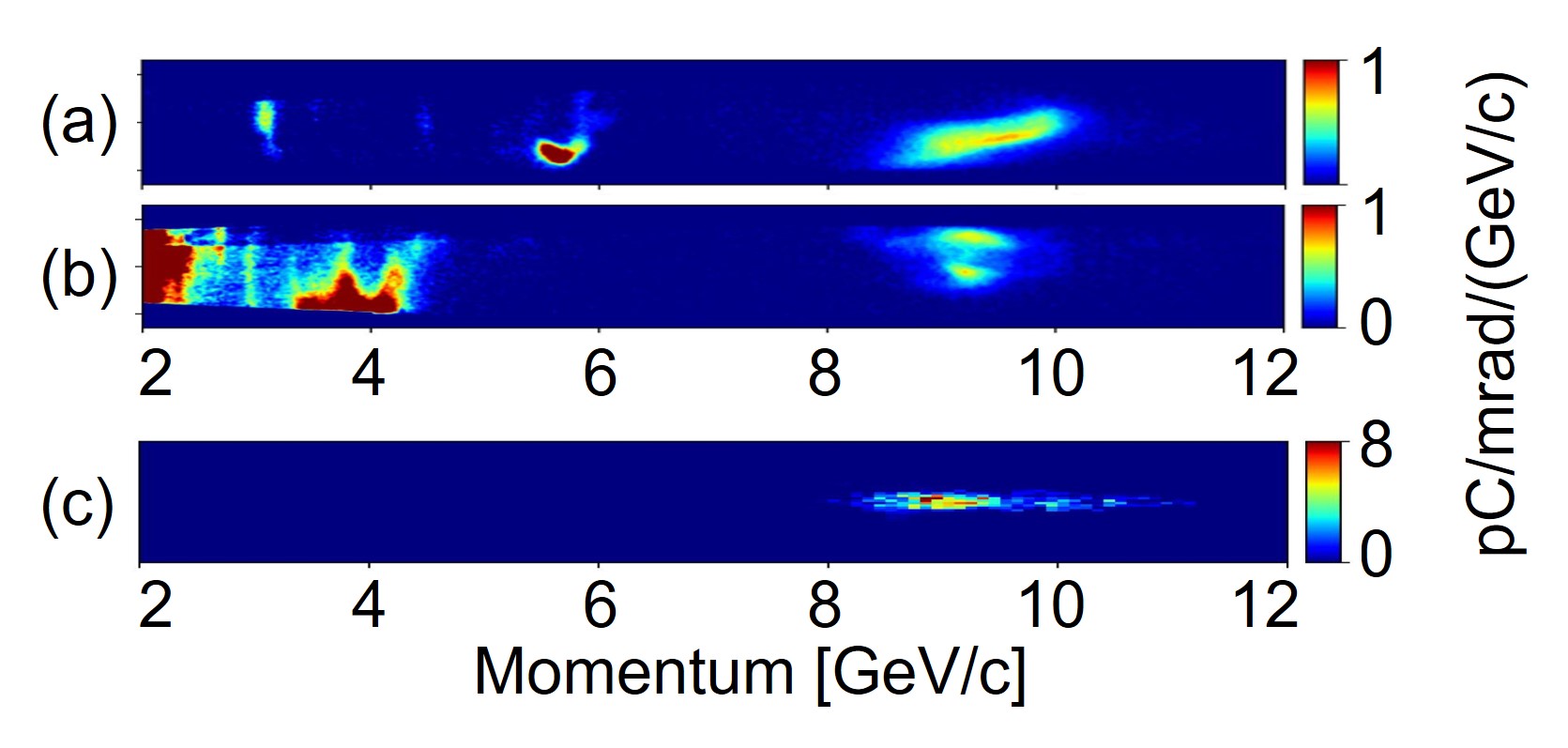}
    \caption{
    (a, b) Examples of measured electron spectra generated in plasma channel using $I_{\mathrm{ch}} = \SI{1.1(1)e15}{W.cm^{-2}}$ and $L_{dop}=20~\rm{cm}$.
    (c) Corresponding simulation result using ($r_{\rm ch}, n_0$) = $( \SI{35}{\micro m}, \SI{6e16}{cm^{-3}} )$.
    }
    \label{fig:fig5}
\end{figure}

We increased $I_{\rm ch}$ to $I_{\mathrm{ch}} = \SI{1.1(1)e15}{W.cm^{-2}}$ (with delay time $\Delta t=5~\rm{ns}$). As shown by the orange curve in fig ~\ref{fig:fig4}, the laser FWHM in-plasma increased to $\approx \SI{45}{\micro m}$ compared to $I_{\mathrm{ch}} = \SI{3.2(4)e14}{W.cm^{-2}}$ where the in-plasma size was $\lesssim \SI{30}{\micro m}$. To observe electron bunch generation, it was necessary to increase the dopant region to $L_{\rm ch}=0-20~\rm{cm}$. Enhanced electron acceleration was observed, as shown in fig \ref{fig:fig5}. Shots that were well aligned to the electron spectrometer indicated a maximum energy of 10.2~GeV. Using the same comparison metric introduced above, the simulation parameters were determined to be ($r_{\rm ch}, ~n_0$) = $( \SI{35}{\micro m}, \SI{6e16}{cm^{-3}} )$. With these parameters, the simulations predict a maximum electron energy of 10.8 GeV at $L_{\rm ch}=40~\rm{cm}$ [Fig.~\ref{fig:fig5}(c)], in good agreement with the experiment. The larger channel radius and higher on-axis density are consistent with hydrodynamic simulations \cite{cook2025hydrodynamic}, which predict that increasing the channel-forming laser intensity produces a wider plasma channel with increased axial density.

Figure \ref{fig:fig6}(a) shows the maximum electron energy for the two cases discussed above. Even though the electrons were injected later for increased $I_{\rm ch}$, increased accelerating gradient of $\sim 30~\rm{GV ~m^{-1}}$ was demonstrated in simulation. This was partially due to the higher on-axis density, but also to transverse effects\cite{McCombs2026inPreparation}. Both the measured and simulated electron beam spectra indicate an approximately linear increase in the maximum electron beam energy, indicating that dephasing was not reached within the maximum length of the accelerator for these conditions. Extending the channel to $\approx \SI{65}{cm}$ would result in electron beams of 10.1 GeV and 15.2 GeV for the low- and high- $I_{\rm ch}$, respectively.

\section{\label{sec:level4}Discussion}

\begin{figure}[t]
    \centering
    \includegraphics[width=1\linewidth]{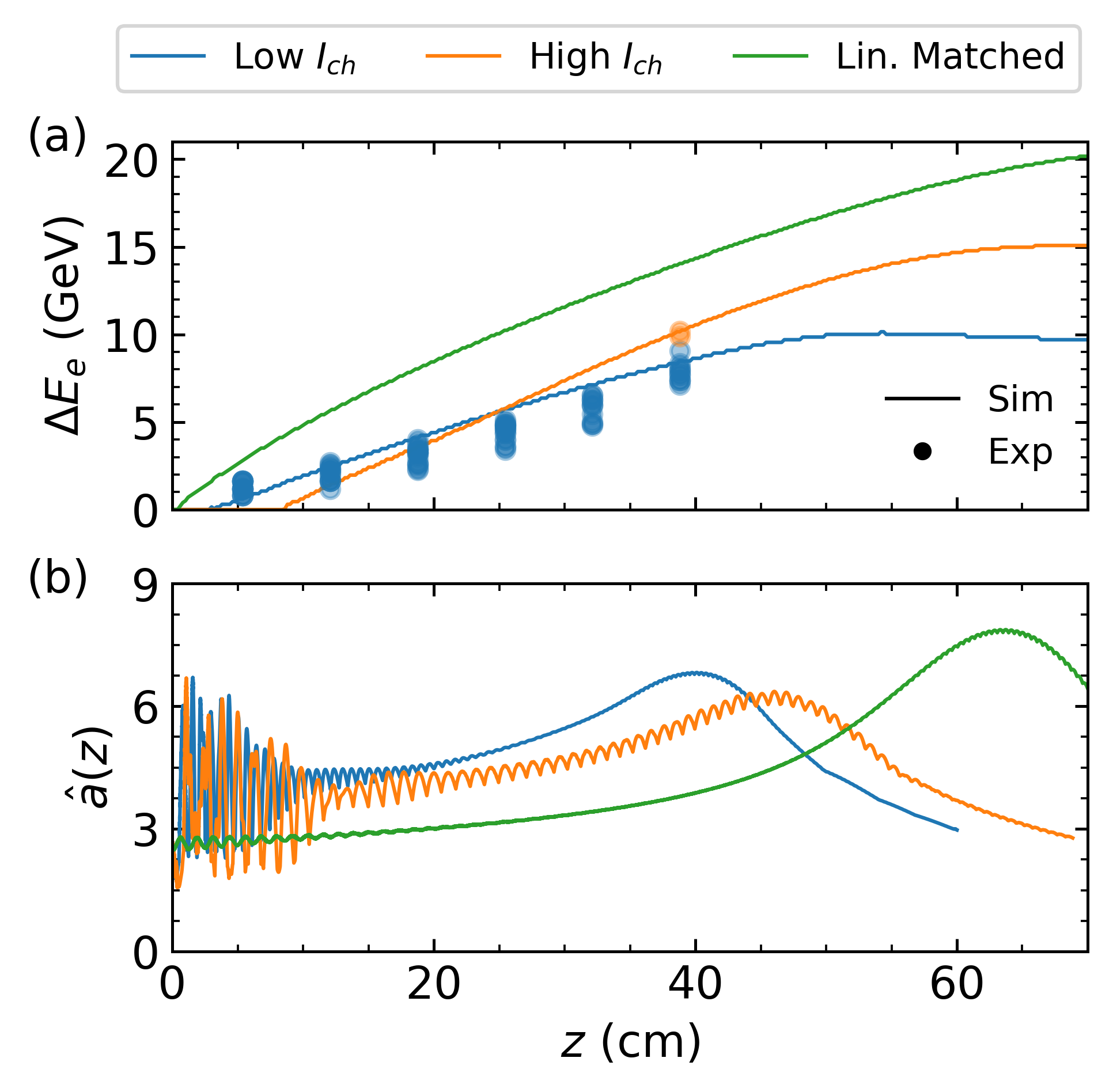}
    \caption{ (a) Maximum electron beam energy as a function of propagation distance. Markers represents one shot well-aligned to the magnetic spectrometer for the conditions of Fig.~\ref{fig:fig1} (blue, $I_{\mathrm{ch}} = \SI{3.2(4)e14}{W.cm^{-2}}$), and Fig.~\ref{fig:fig5} (orange, $I_{\mathrm{ch}} = \SI{1.1(1)e15}{W.cm^{-2}}$). Corresponding solid lines indicated the simulation maximum energy. (b) Calculated normalized vector potential of the driver laser as a function of propagation distance for the two $I_{\rm ch}$ and experimentally measured drive laser. The green curve shows the case where the drive laser transverse mode is linearly matched to the high-$I$ channel, for a laser energy of 24 J, a pulse duration of 80 fs.
    }
    \label{fig:fig6}
\end{figure}

With all simulation and experimental observables in excellent agreement, the simulations can now be used to investigate the physics of the interaction not directly observable in the experiment, and understand details of channel-guided accelerators. Figure \ref{fig:fig6}(b) shows the calculated evolution of the normalized peak laser intensity $\hat{a}(z)$ for the simulations that meet the constrained parameters at low $I_{\mathrm{ch}} = \SI{3.2(4)e14}{W.cm^{-2}}$ and increased $I_{\mathrm{ch}} = \SI{1.1(1)e15}{W.cm^{-2}}$. Periodic oscillations in $\hat{a}(z)$ exhibited in both cases can be explained by beating of higher-order modes \cite{esarey1998nonparaxial, shrock2024guided, picksley2024prl}, which originate in the following way. For a parabolic plasma channel, the solutions to the Helmholtz equation are a set of Laguerre-Gauss transverse modes $E_{pm}$ where $(p,m)$ refer to the radial and azimuthal mode number respectively. The lowest order is a Gaussian $E_{00} \propto \exp[-(r/r_{\rm ch})^2]$, and hence a low-power Gaussian transverse mode with $w_0 = r_{\rm ch}$ propagates with constant spot-size and transverse shape. HOFI plasma channels are not parabolic and extend only to a finite radius (the neutral collar surrounding the initial shock), hence their mode structure must be calculated numerically from $n(r)$ \cite{clark2000optical}. The resulting fundamental channel mode is a perturbation from Gaussian, which we refer to as the linearly matched mode.

Since the laser focal profile is not linearly matched, one can express the laser field $E_\mathrm{L}(r)$ as a linear combination of several channel modes, including the fundamental and higher-order modes. Since HOFI plasma channels are finite in extent, only a few low-order modes are allowed to propagate. The leakage rate of each mode increases with $p$ and $m$ respectively \cite{clark2000optical}, and the group velocity of higher order modes is lower than that of the fundamental (modal dispersion) \cite{esarey1998nonparaxial, schroeder2011group, shrock2024guided}. Hence, maximal transfer of laser energy to the plasma wave occurs when the laser is matched to the fundamental mode of the plasma channel.

The periodic oscillations in laser intensity can result in a transverse wakefield structure that is unsuitable for the transport of electron beams \cite{picksley2023all, shrock2024guided, picksley2024prl}. This is directly observable by comparing the two different $I_{\rm ch}$ in fig \ref{fig:fig6}. At low $I_{\rm ch}$, where $r_\mathrm{ch}$ was smaller, $\hat{a}$ enhancement for $z \lesssim \SI{7}{cm}$ is greater due to greater focusing by the channel, and the distance over which higher-order modes were separated was shorter (modal dispersion is greater). Ionization injection is promoted for $z \lesssim \SI{3}{cm}$, as we explicitly demonstrated in our experiments [see Fig.~\ref{fig:fig1}(c)]. This in turn resulted in reduced coupling of laser energy into the plasma channel, and a lower accelerating gradient, owing to the reduced wake size in a narrow channel. 
For increased $I_{\rm ch}$ and hence $r_\mathrm{ch}$, modal dispersion occurs over a longer length, and injection is delayed to $z \approx \SI{9.5}{cm}$, similar to our previous work described in Ref.~\cite{picksley2024prl}. However, the larger $r_\mathrm{ch}$ produces a larger wake, which—together with the higher density—results in a greater accelerating gradient.

It is useful to compare the propagation of the experimentally measured laser pulse with the case where the transverse profile of the laser is linearly matched to the fundamental mode of the plasma channel [green lines in Fig ~\ref{fig:fig6}]. Mode-beating oscillations in $\hat{a}$ are greatly reduced in the case of linear matching, and ionization injection is promoted from $z \approx 0$ for suitable values of laser energy, 24 J, and duration 80 fs (values optimized for electron beam trapping). Hence, for the same laser energy, and plasma channel used in this experiment, the linearly matched case reached a simulated energy gain of $\SI{20.2}{GeV}$ after an acceleration distance of $\SI{69}{cm}$. The higher energy, relative to the simulation using the measured transverse mode, arises because injection occurs at $z \approx 0$; the largely reduced mode beating also permits a longer drive pulse, which is better matched to the plasma resonance.

These results demonstrate how matching of the plasma channel shape to the laser can significantly affect the overall dynamics of acceleration. For current state-of-the-art, PW-class Ti:Sapphire lasers, the laser transverse (and longitudinal) shape cannot be perfectly matched to the plasma channel itself. Instead, the plasma channel radius can be adjusted only to maximize coupling of laser energy into the fundamental channel mode.

\section{\label{sec:level5}Conclusion}

This work addresses a central limitation in interpreting channel-guided laser–plasma acceleration experiments: substantial uncertainty in the transverse plasma density profile can lead to strong degeneracy in simulations, where distinct channel structures produce similar electron-beam outcomes when only exit-plane measurements are considered. By combining longitudinally resolved measurements of electron-beam evolution with independent diagnostics of laser spectral evolution, we show that these observables must be analyzed jointly to constrain the experimental plasma channel parameters and the underlying interaction physics.
Using a parameter scan performed with PIC simulations, we demonstrate that laser redshift alone does not uniquely determine the channel properties. Introducing a combined metric based on agreement in laser redshift and electron-beam energy evolution identifies a single parameter set that simultaneously reproduces the measured laser evolution and electron acceleration, thereby constraining the transverse channel profile relevant for guiding. The constrained modeling indicates that the initial configuration corresponds to a mismatched guiding condition, motivating targeted optimization of the channel formation.
We then demonstrate such optimization experimentally by increasing the channel-forming laser intensity, which produces improved guiding consistent with a wider channel and higher on-axis density. Under these conditions, electron energies exceeding 10 GeV are achieved in a 40 cm channel. Finally, we show that plasma density down-ramps significantly affect inference of the in-plasma laser spot size from downstream measurements, underscoring the importance of including ramps when using external diagnostics to validate simulations. Together, these results establish a practical, multi-observable strategy for reducing parameter uncertainty in channel-guided LPAs and for using constrained modeling to guide accelerator optimization toward application-relevant beams.

\begin{acknowledgments}
This work was supported by the Director, Office of Science, Office of High Energy Physics, of the U.S. Department of Energy under Contract No. DE-AC02-05CH11231, and used the facilities at the National Energy Research Scientific Computing Center (NERSC) under award HEP-ERCAP0035612. We greatly acknowledge technical support from Zac Eisentraut, Teo Maldonado Mancuso, Chetanya Jain, Mark Kirkpatrick, Federico Mazzini, Nathan Ybarrolaza, Derrick McGrew, Paul Centeno, Art Magana, Mackinley Kath, and Joe Riley. The authors would like to thank Nathan Cook, Remi Lehe, and Aodhan McIlvenny for useful discussions. 
\end{acknowledgments}


\bibliography{apssamp}

\end{document}